\documentclass[superscriptaddress]{revtex4}

\usepackage{graphicx}   
\usepackage{color}
\usepackage[10pt]{moresize}
\usepackage{bm}
\usepackage{amsfonts}
\usepackage{amssymb}
\usepackage{amscd}
\usepackage{amsmath}    
\usepackage{enumerate}
\usepackage{epsfig}
\usepackage{subfigure}
\usepackage{color}
\usepackage{amsthm}

\begin{document}

\title{Random number generation with cosmic photons}

\author{Cheng Wu}
\author{Bing Bai}
\author{Yang Liu}
\affiliation{Shanghai Branch, Department of Modern Physics and National Laboratory for Physical Sciences at the Microscale, University of Science and Technology of China, Shanghai 201315, China}
\affiliation{CAS Center for Excellence and Synergetic Innovation Center in Quantum Information and Quantum Physics, University of Science and Technology of China, Shanghai 201315, China}

\author{Xiaoming Zhang}
\affiliation{Key Laboratory of Optical Astronomy, National Astronomical Observatories, Chinese Academy of Sciences, Beijing 100012, China}

\author{Meng Yang}
\author{Yuan Cao}
\affiliation{Shanghai Branch, Department of Modern Physics and National Laboratory for Physical Sciences at the Microscale, University of Science and Technology of China, Shanghai 201315, China}
\affiliation{CAS Center for Excellence and Synergetic Innovation Center in Quantum Information and Quantum Physics, University of Science and Technology of China, Shanghai 201315, China}

\author{Jianfeng Wang}
\affiliation{Key Laboratory of Optical Astronomy, National Astronomical Observatories, Chinese Academy of Sciences, Beijing 100012, China}

\author{Shaohua Zhang}
\author{Hongyan Zhou}
\author{Xiheng Shi}
\affiliation{Polar Research Institute of China, Shanghai 200136, China}

\author{Xiongfeng Ma}
\affiliation{Center for Quantum Information, Institute for Interdisciplinary Information Sciences, Tsinghua University, Beijing 100084, China}

\author{Ji-Gang Ren}
\author{Jun Zhang}
\author{Cheng-Zhi Peng}
\affiliation{Shanghai Branch, Department of Modern Physics and National Laboratory for Physical Sciences at the Microscale, University of Science and Technology of China, Shanghai 201315, China}
\affiliation{CAS Center for Excellence and Synergetic Innovation Center in Quantum Information and Quantum Physics, University of Science and Technology of China, Shanghai 201315, China}

\author{Jingyun Fan}
\email{fanjy@ustc.edu.cn}
\author{Qiang Zhang}
\email{qiangzh@ustc.edu.cn}
\author{Jian-Wei Pan}
\email{pan@ustc.edu.cn}
\affiliation{Shanghai Branch, Department of Modern Physics and National Laboratory for Physical Sciences at the Microscale, University of Science and Technology of China, Shanghai 201315, China}
\affiliation{CAS Center for Excellence and Synergetic Innovation Center in Quantum Information and Quantum Physics, University of Science and Technology of China, Shanghai 201315, China}

\begin{abstract}
Random numbers are indispensable for a variety of applications ranging from testing physics foundation to information encryption. In particular, nonlocality tests provide a strong evidence to our current understanding of nature --- quantum mechanics. All the random number generators (RNG) used for the existing tests are constructed locally, making the test results vulnerable to the freedom-of-choice loophole. We report an experimental realization of RNGs based on the arrival time of cosmic photons. The measurement outcomes (raw data) pass the standard NIST statistical test suite. We present a realistic design to employ these RNGs in a Bell test experiment, which addresses the freedom-of-choice loophole.

\end{abstract}

\maketitle

{\it Introduction.---}
Randomness is one of the most fundamental features of nature. The best example may be the biological diversity \cite{hartwell1999molecular}. Another example is Brownian motion \cite{kac1947random,grabert1983quantum} which has been studied for nearly two centuries. Random number generators (RNG) are based on either a classical mechanism or a quantum process. Quantum random number generators (QRNGs) rely on breaking quantum superpositions results into unpredictable measurement outcome and are therefore deemed to be truly random. A number of quantum processes are utilized to make QRNGs (for reviews see Ref. \cite{MaQRNG,herrero2016quantum} and references therein). 

Bell tests, or experimental violation of Bell's inequality, provide a strong support to quantum mechanics, especially to rule out local hidden variable models. Recently, both locality and efficiency loopholes were closed in Bell test experiments \cite{hensen2015loophole,shalm2015strong,giustina2015significant}, in which QRNGs were employed in state measurements at two remote test sites. However, the test results may not be reliable if the two RNGs are somehow correlated (with each other and/or with the two physical devices). For instance, the distant entangled photon pairs in a Bell test learn the random inputs before they are separated. This is called freedom-of-choice loophole (also known as randomness loophole). The time constraint for a local hidden variable mechanism to occur to affect the test results in previous loophole free Bell experiments is less than $10^{-5}$ s before the experiment, which may be pushed deep into the cosmic history by adopting the RNG scheme based on cosmic photon measurements to take advantage of randomness at remote celestial objects, e.g., measuring the temporal mode of photons as studied in this paper.  The randomness of the outcomes cannot be proven strictly, but it is supported by following physical observations. First, the setup measures the arrival time of photons from the celestial object which the telescope points at. Second, the generation time of cosmic photons from a celestial object is random, so does the arrival time. The states of photons from two celestial objects are independent. This is related to the no-signaling assumption. In a way, we assume that the nature is not malicious to jeopardize our experiment.
We realize RNGs with photons from an array of cosmic radiation sources with magnitude between 4.85 and 13.5 and distance (from Earth) between 756 and $7.49 \times 10^8$ light years (ly). These RNGs can deliver raw random bits exceeding $10^{6}s^{-1}$, which pass NIST statistical test suite. We present a realistic design of event-ready Bell test experiment with these RNGs to address the freedom-of-choice loophole while closing locality and efficiency loopholes simultaneously.

\begin{figure}[tbh]
\centering
\includegraphics[width=8cm,height=5cm]{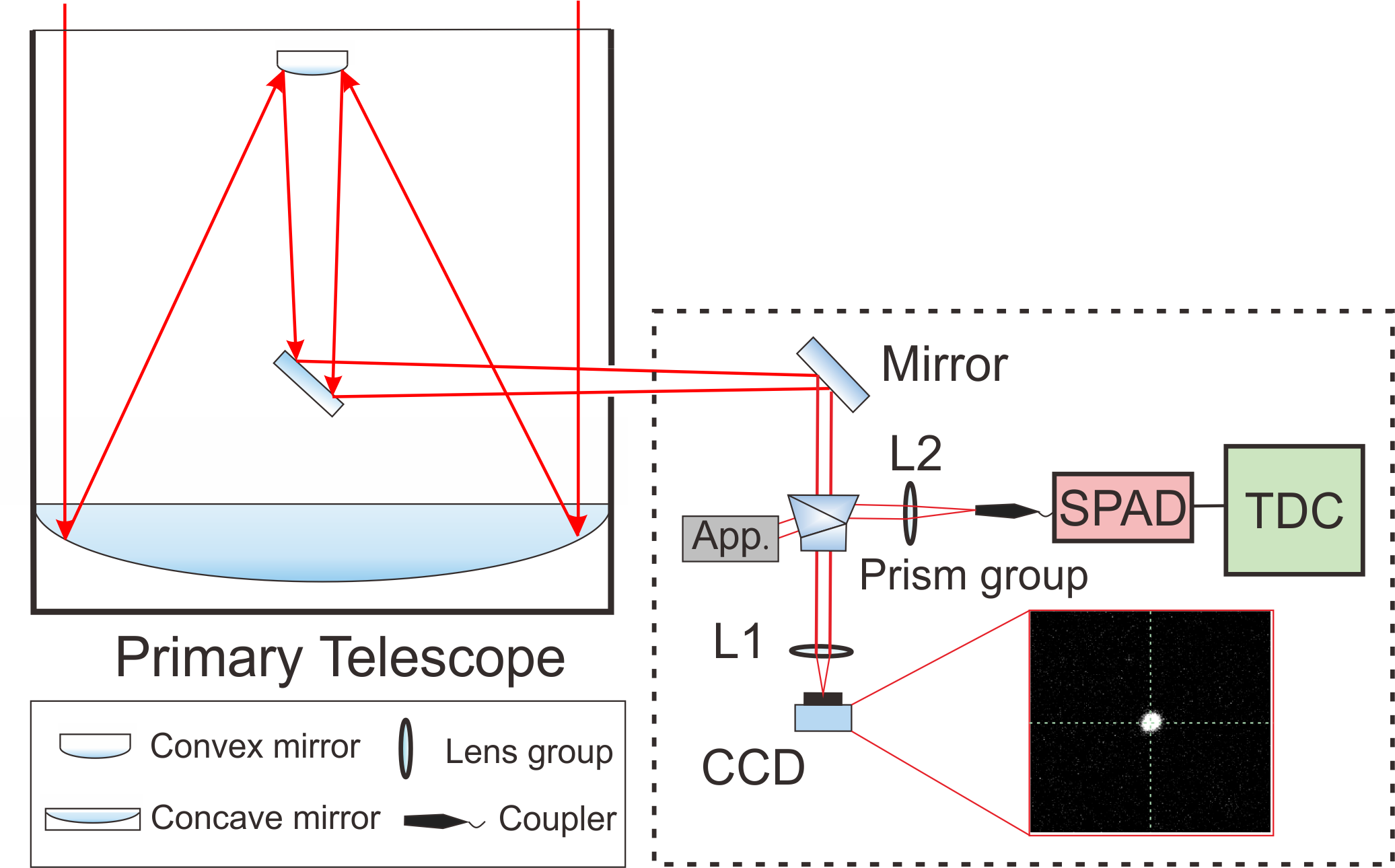}
\caption{Random number generation with cosmic photons. Photons from a cosmic radiation source under study (CRSS) with wavelength in the range, [680, 830] nm are collected into a multimode optical fiber for RNG. Inset: Photons with wavelength in the range, [530, 680] nm, form an image of the CRSS, here, quasar IGR J03334+371 on the camera for tracking. APP: astronomy applications.}
\label{setup}
\end{figure}

\begin{table*}[htb]
\caption{Photon counting data for cosmic radiation sources under study (see SM about distance \cite{SM-ref}) \cite{HIP,masetti2008unveiling,astraatmadja2016estimating}}
\centering
\begin{tabular}{cccccccc}
\hline\hline
Name & Magnitude & Distance& Signal rate& total Data& background & r & min-entropy \\
 & &(ly)&($\times 10^6 s^{-1}$)&(Gb)&($s^{-1}$)&& H\\
\hline
HIP15416 \cite{HIP}&	4.85&	1177&	2.20-2.28&	1&	914&	2450&	0.9969\\
HIP117447 \cite{HIP}&	5.43&	6151&	0.86-1.2&	1&	512&	2012&	0.9978\\
HIP2876 \cite{astraatmadja2016estimating}&	5.75&	2675&	0.51-0.53&	1&	464&	1130&	0.9981\\
HIP6522 \cite{astraatmadja2016estimating}&	6.07&	5488&	0.48-0.68&	1&	578&	1010&	0.9983\\
HIP3030 \cite{HIP}&	6.75&	5344&	0.64-0.65&	1&	518&	1260&	0.9976\\
HIP100548 \cite{HIP}&	7.03&	5621&	0.33-0.52&	1&	615&	680&	0.9973\\
HD33339 \cite{astraatmadja2016estimating}&	7.99&	756&0.23-0.24&	1&	662&	350&	0.9980\\
HIP20276 \cite{astraatmadja2016estimating}&	8.24&	1835&	0.18-0.26&	1&	486&	400&	0.9980\\
HIP3752 \cite{astraatmadja2016estimating}&	9.02&	908&	0.12-0.13&	1&	532&	235&	0.9973\\
HIP114579 \cite{astraatmadja2016estimating}&	9.27&	1967&	0.05-0.10&	1&	442&	170&	0.9974\\
HIP117690 \cite{astraatmadja2016estimating}&	9.9&	21733&	0.033-0.043&	1&	674&	57&	0.9938\\
HIP23114 \cite{astraatmadja2016estimating}&	10.6&	2243&	0.009-0.013&	0.1&	417&	28&	0.9909\\
IGR J03334+371 \cite{masetti2008unveiling}&	13.5&7.49 $\times$ $10^8$	&0.0011-0.0031&	0.1&	359&	8		&0.9897\\
\hline
\hline
\end{tabular}
\label{tab:divin}
\end{table*}

{\it Random number generation with cosmic photons.---}
The experiment is conducted in the Astronomy Observatory at Xinglong, China (N $40^{\circ}23.75'$, E $117^{\circ}34.5'$). We use a Ritchey-Chretien (RC) optical telescope with a diameter of 1 meter and a focal length of $f = 5 $ meter to collect light from the cosmic radiation source under study (CRSS) and use prisms to direct lights of various spectral bands to different applications. The light that is incident onto this RC telescope from a typical cosmic radiation source with an angular spread of $\phi = 3''$ has an 1/e-diameter of 73 $\mu m$ and a numerical aperture (NA) of 0.10 at the focal plane. A multimode optical fiber with NA = 0.22 and a core diameter of 105 $\mu m$ is placed at the focal plane to collect light with wavelength in the range, [680, 830] nm, and direct the light to a single photon avalanche diode (SPAD, model: EXCELITAS, active area: 170 $\mu m$, single photon detection efficiency: $\sim 55 \%$ at 780 nm). A CCD camera is also placed at the focal plane to image the CRSS by detecting light with wavelength in the range, [530, 680] nm. We stabilize the coupling of cosmic photons from the CRSS into the multimode fiber by a standard altitude-azimuth tracking mechanism \cite{kwok2016improving}. We estimate the total detection efficiency of single cosmic photon to be about 2\% (see Supplemental Material (SM) \cite{SM-ref}). So we require an assumption that the detected photons represent a fair sample of photons emitted by the CRSS. In addition, as we assume in the above that the nature does not maliciously jeopardize our experiment, we assume that the propagation of cosmic photons and their arrival times are not affected by any mechanism other than the known mechanisms in astronomy studies such as refraction through slowly varying interstellar and intergalactic media and assume that the effect is identical for all photons \cite{gallicchio2014testing}. In fact, there was no astronomy report on delaying the cosmic photon arrival time at the visible or near infrared wavelength \cite{cordes2002}. We consider the major delay may be due to the refractive index of atmosphere around us and include it in the discussion below.

We choose to detect cosmic photons over a bandwidth of 150 nm to increase the rate of random bits, which features the Poissonian statistics: the mean photon number is a constant for equal time and the time interval between photon emission events is random. If the period $T_W$ of a reference clock is equally divided into N time bins, the probability for a cosmic photon to arrive at an arbitrary time bin $t_i$ (i = 1, 2, $\ldots$, N) is a constant, $P_i = 1 / N$. In our experiment, the photon-detection signal from the SPAD is recorded using a home-made time-to-digital converter (TDC) with a time resolution of 25 ps. We set $T_W$ = 40.96 ns to be smaller than the recovery time (45 ns) of the SPAD such that there is at most one detection event per clock cycle, and set N = 256 ($\times$ 160 ps). We assign each time bin with an unique 8-bit binary code, and record the assigned code for the time bin in which a detection event occurs.

We study RNGs with photons from an array of celestial objects. The main results are summarized in Table I. First, we notice that the photon counting signal rates exceed $10^{6}s^{-1}$ (which is within the linear operation mode of the SPAD in use) for CRSS with lower magnitude, demonstrating that this method is as efficient as laser-based RNGs in generating random numbers \cite{jennewein2000fast, Gabriel2010generator, symul2011real, nie2014practical, abellan2014ultra, applegate2015efficient, abellan2015generation}. Second, despite the dramatic fluctuation of signal rates (shown by signal ranges in Table I ), the true signal rate (with background subtracted) scales with magnitude as expected, as indicated in Fig. 2. We attribute fluctuation in the rate partly to atmospheric disturbance. Besides attenuating cosmic light, atmospheric disturbances deteriorate the coupling of light from CRSS into the multimode fiber by either displacing focal position or modifying beam profile, which can not be corrected by current tracking mechanism. We use the maximum rate for each CRSS to suppress such impact in generating the trend line. ( We do not use the data for quasar IGR J03334+371 (red filled dots) in generating the trend line because of small signal-to-noise ratio.)

\begin{figure}[tbh]
\centering
\includegraphics[width=8cm,height=5cm]{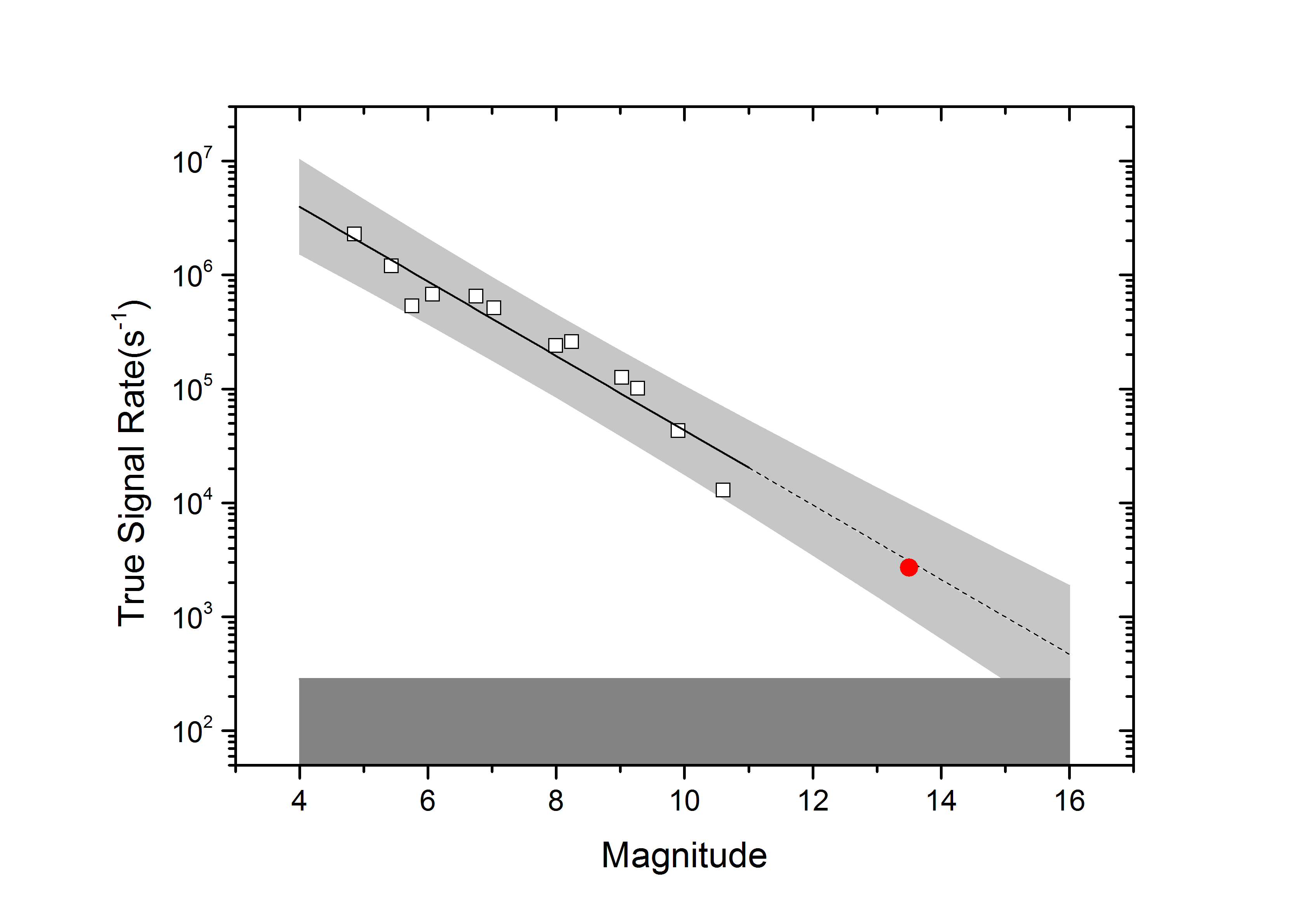}
\caption{Experimental true signal rate (background subtracted) versus magnitude. The trend line (solid) is fitted with data (open square) for magnitude $< 11$, and extrapolated to magnitude 16 (dashed line). The shaded regions indicate 2-standard deviations, with the one on horizontal axis for background. Red filled dot: data for quasar IGR J03334+371.}
\label{}
\end{figure}


We use raw data in the analysis. For each CRSS, the probability of photon arrival time ($P_i$) is uniformly distributed around the ideal value of 1/256, indicating a good level of randomness (see SM \cite{SM-ref}). We apply two standard methods to evaluate the performance of cosmic photon RNGs. The min-entropy, $H_\infty =-\log (\max {P_i})$, is consistent with the ideal value of one within 1\%, and the raw data pass the NIST statistical test suite \cite{nist2010special} (see SM \cite{SM-ref}).
These two results certify the quality of these cosmic RNGs. 

To estimate the background contribution, we point the RC telescope slightly away from the CRSS (at a dark patch of the sky) till the detection rate drops to a stable level. The background may include contributions due to detector dark counts or ground-based light sources, which can be used by local hidden variable theories and must be made insignificant. Below we present a realistic analysis on Bell experiment with cosmic RNGs with large signal-to-noise ratio.

{\it Event-ready Bell test experiment with cosmic RNGs---}
The celebrated Bell's inequality \cite{bell1964einstein,bell1987speakable} is based on the assumption of locality, realism and freedom-of-choice. In previous Bell test experiments \cite{hensen2015loophole,giustina2015significant,shalm2015strong} with a pair of entangled particles A and B, the events for entanglement generation, base choices, and state measurements are separated space-like in future light cones. However, these light cones cross each other in $< 10^{-5}$ s in the past direction, allowing the possibility for local correlation events occurring in the overlapped regions to control measurement outcomes. Furthermore, it was shown that a Bell test experiment is vulnerable to local hidden variable theories even with a conspiracy of as little as 1/22 bit of mutual information between RNGs and source of entanglement \cite{hall2011relaxed}. Here we consider two possible scenarios that local correlation events may impact the experimental outcomes as shown in Fig. 3. In the first case, a local correlation event Y1 (Y2) may share information, denoted by a local hidden variable $\lambda1$($\lambda2$), about photon emission event S1 (S2) for random bit generation with the source, prior to state preparation, provided that local correlations take place ahead at least by an amount of time, $\tau_1 \geq \min(L_{1}/c,L_{2}/c)$, where $L_1,L_2$ are distances of the two cosmic sources from Earth.   In the second case, a local correlation event (denoted by a hidden variable $\lambda3$) may occur in the overlapped region formed by the past light cones of two cosmic photon radiation events, S1 and S2, prior to the experiment by  $\tau_2 \geq (L_1 + L_2 + L_{12}) / 2c \geq \tau_1$, where $L_{12}$ is distance between the two cosmic sources (see SM \cite{SM-ref}). Therefore, local correlation events in the green shaded regions may impact the outcomes of Bell test experiment as shown in Fig. 3, with the time constraint to be $\tau \ge \tau_1$. For example, by employing RNGs based on cosmic sources HIP 55892 ($L_1 = 3325 \pm 1649 $ lys, Magnitude 6.7) and and HIP 117928  ($L_2 = 3454 \pm 1433 $ lys, Magnitude 8.9) \cite{astraatmadja2016estimating}, we have $\tau \ge 3325 \pm 1649 $ years (see SM \cite{SM-ref}), which is $\sim$ 16 orders of magnitude improvement over previous loophole free Bell test experiments.


\begin{figure}[tbh]
\centering
\includegraphics [width=8cm,height=6cm]{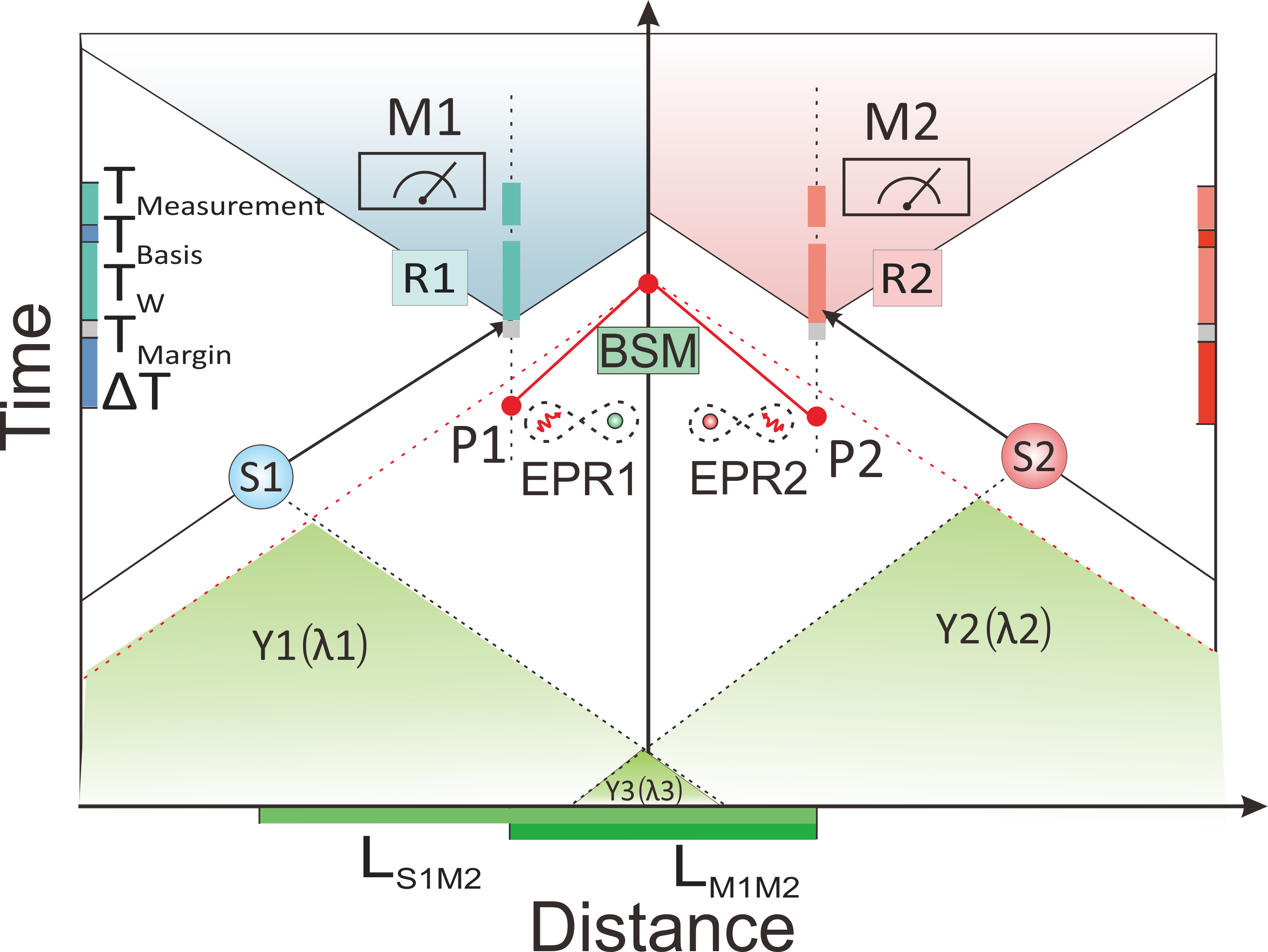}
\caption{Space-time diagram of an event-ready Bell test experiment with NV-centers \cite{hensen2015loophole,gallicchio2014testing}. S1 and S2 are cosmic photon emission events, followed by events R1 and R2 to output random bits for base choice. P1 and P2 are events for NV centers to send photons for Bell state measurement (BSM) after the creation of entangled photon-electron pairs (EPR1,2) at NV centers. The photons are sent for BSM via optical fibers, shown by red lines.  A destructive BSM with photons from the two entangled electron-photon pairs prepares the two electrons in a Bell state, which is ready for state measurements (M1 and M2) after the base choice. Y1, Y2 and Y3 are local correlation events (denoted by local hidden variables, $\lambda1$, $\lambda2$, $\lambda3$) that may occur in the overlapped region formed by the past light cones (see text for details). }
\label{fig:capacity}
\end{figure}

\begin{figure}[tbh]
\centering
\includegraphics [width=8cm,height=5cm]{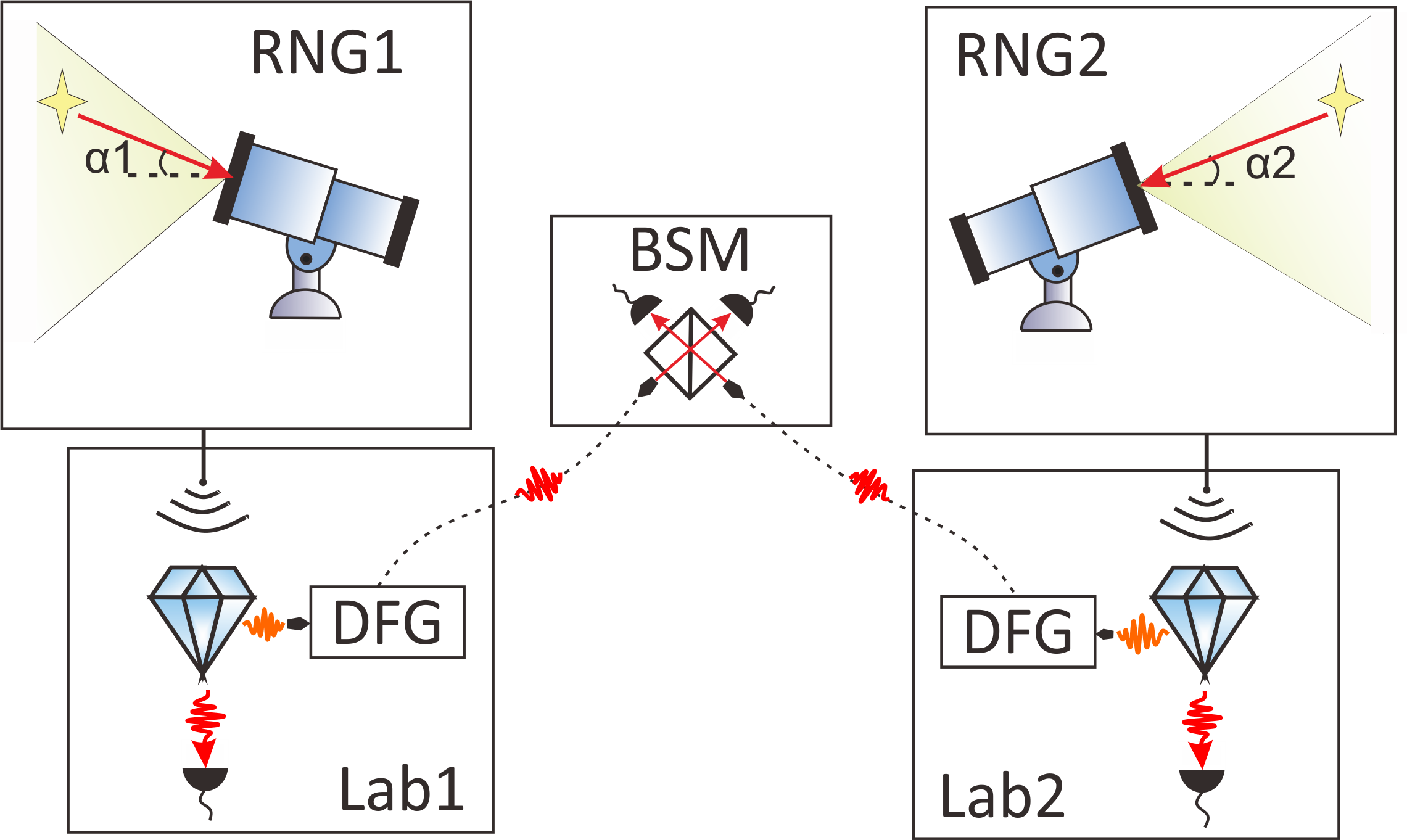}
\caption{Schematic of an event-ready Bell experiment. Each measurement station (Lab1 or Lab2) prepares an entangled electron-photon pair with a NV center. Lab1(2) downconverts single photons from visible to infrared (1550 nm) via difference frequency generation (DFG). A successful BSM with a single photon from Lab1 and a single photon from Lab2 projects the corresponding two electrons into a Bell state. RNG1 and RNG2 provide random bits per time window $T_W$ to set base in measuring the quantum state of electron spin. $\alpha1$ and $\alpha2$ are angles of the optical axes of telescopes with respect to the Lab axis (see SM \cite{SM-ref}).}
\label{}
\end{figure}
			
Below we present a realistic design to use the two RNGs (with signal-to-noise ratio $>$ 100) in an event-ready Bell test experiment with NV centers \cite{hensen2015loophole}, as shown in Fig. 4.

First, locality requests space-like separation between event of state measurement and event of base choice, which requires that the distance $L_{M1M2}$ between two measurement stations Lab1 and Lab2 is set according to $T_{W} + T_{Basis} + T_{Measurement} + T_{Margin} < L_{M1M2} / c \cdot \cos \alpha$, where $T_{Basis}$ is the time elapsed for event completing the base choice upon receiving a random bit, $T_{Measurement}$ is the time elapsed for event completing the state measurement after the base choice, $T_{Margin}$ accounts for possible additional delays and $\alpha$ is elevation angle of telescope (see SM \cite{SM-ref}).  Taking $T_{W} = $ 8 $\mu s$, $T_{Measurement} \sim$ 4 $\mu s$, $T_{Basis} \sim $ 1 $\mu s$, $T_{Margin} \sim $ 1 $ \mu s$ and  $\alpha \leq 30^{\circ}$ for telescopes, we have $L_{M1M2} >$ 5 km. Second, consideration of freedom-of-choice requests space-like separation between event of cosmic photon emission and event of two-photon Bell state measurement (BSM), so photon-electron entanglement must be produced in NV-center in advance by $\Delta T > (1.45-\cos \alpha) \cdot L_{M1M2} / 2c \sim$ 5 $\mu s$, which is much shorter than the coherence time (T2 $\sim 0.6 s$ at 77 K \cite{bar2013solid}) of electron spin of NV center (see SM \cite{SM-ref}).

Photons emitted by NV centers at the visible wavelength ($\sim$ 640 nm) are downconverted to photons at the wavelength of $\sim$ 1550 nm via difference frequency generation (DFG), because they are subject to high propagation loss in optical fiber. Considering 1.5 dB loss due to downconversion operation \cite{de2012quantum} and 1 dB loss due to photon propagation over the 5 km optical fiber, there is an improvement of $\sim$ 10 dB in two-photon detection efficiency over the previous experiment \cite{hensen2015loophole}. With that, the averaged success probability per entanglement generation attempt is estimated to be $P_{total} \sim 2.26 \times 10^{-9}$ (see SM \cite{SM-ref}). This will result in one event-ready electron pair entanglement per 0.29 hours per 24$\mu s$ measurement period on average. So it will take about 72 hours to violate the Bell's inequality with statistical confidence similar to the previous experiment \cite{hensen2015loophole}. More importantly, the time constraint for local hidden variable mechanism to impact the outcome of Bell test experiment is moved by more than 1000 years back into the past.

{\it Discussions.---}
It was recently proposed to push the time constraint to reject local hidden variable mechanisms in a Bell test experiment by billions of years back into the cosmic history by employing RNGs with photons from quasars of high redshift \cite{saturni2016multi}. We discuss about its practicability by analyzing the performance of a RNG with photons from quasar APM 08279+5255 with magnitude 15.3 and redshift z = 3.91. According to the trend line in Fig. 2, the true signal rate of this RNG is $\sim$ 590 $ s^{-1}$ (at $\alpha = 30^\circ$). We attribute the low signal-to-noise ratio $\sim$ 2 (for background rate 550 $s^{-1}$) mainly to that the optical system is not operated optimally. The signal-to-noise ratio can be increased to $>$ 50 by having the RC-telescope work in the diffraction limit \cite{wizinowich2000first}, and $>$ 100 in the space due to reduced sky brightness and absence of atmosphere attenuation to cosmic photons. The absence of atmospheric disturbance and angular separation of $180^{\circ}$ between two telescopes are also advantages of a satellite-based cosmic Bell experiment (see SM \cite{SM-ref}).


{\it Conclusion.---}
In conclusion, we realize cosmic-photon base RNGs and present a realistic design to use these RNGs in a Bell test experiment. We show that it is experimentally feasible to perform a Bell test experiment with RNGs based on quasars of high redshift, which will provide a strong support to quantum mechanics, by setting the time constraint to reject local hidden variable mechanisms deep into the cosmic history. Meanwhile, the method of single-photon detection of cosmic photons may provide a powerful tool for cosmology observation.

{\it Acknowledgments.---}
The authors would like to thank Z.-P.~Li, D.-D.~Li, X.~Han and X.~Pan for technical assistance. We acknowledge the support of the staff of the Xinglong 1m telescope. This work has been supported by the National Fundamental Research Program Grant No.~2013CB336800, the Chinese Academy of Science, the National Natural Science Foundation of China, and the Open Project Program of the Key Laboratory of Optical Astronomy, National Astronomical Observatories.

C.~W. and B.~B. contributed equally to this work.

Note added: We became aware of a relevant work during the submission \cite{Johannes2017}.

%

\bibliographystyle{apsrev4-1}

\end{document}


\title{Supplemental Materials: Random Number Generation with Cosmic Photons}

\maketitle

\section{Distance between cosmic source and Earth}
Distance estimation of celestial objects from Earth remains a technical challenge. For example, the distance for HIP 114579 was reported to be 163,000 lys in year 1997, but was updated to be 1967 lys in year 2016.
Table 1 shows the changes in distance estimation of celestial objects (used in our study) over the past years \cite{ perryman1997hipparcos, van2007validation, astraatmadja2016estimating}, which was mainly based on mission Hipparcos (https://www.cosmos.esa.int/web/hipparcos) and mission Gaia (http://sci.esa.int/gaia/) of European Space Agency. For each celestial object, we used the data from the latest astronomy observations. For example, for HIP 2876, the latest data was released in year 2016; for HIP15416, the latest data was released in year 2007. The relative uncertainty in distance estimation is typically on the order of 50\% or higher for distances estimated to be more than 1000 lys.
For the event-ready Bell experiment discussed in the main text, we select two celestial objects, HIP 55892 and HIP 117928 for RNGs. The distance of HIP 55892 from Earth was estimated to be 3881 lys in 2007 and 3325 lys in 2016 (with a relative change of 14\%) and the distance of HIP 117928 from Earth was estimated to be 3835 lys in 2007 and 3454 lys in 2016 (with a relative difference of 10\%). Because the variations are relatively small between the 2007 data and the 2016 data and cosmic RNGs with the two cosmic sources meet the space-like separation requirement in the proposed Bell experiment, we choose the two cosmic sources in our discussion in the main text.

\begin{table*}[htb]
\begin{threeparttable}
\caption{Distance of  cosmic radiation source under study (CRSS) from Earth\tnote{1}.}
    \centering
    \begin{tabular}{cccccc}
    \hline\hline
Year&	1997 \cite{perryman1997hipparcos}&	\multicolumn {2}{c} {2007 \cite{van2007validation}}&	\multicolumn {2}{c} {2016 \cite{astraatmadja2016estimating}}\\
CSSR&	Distance (lys)&	Distance (lys)&	$\sigma$ (lys)&Distance(lys)&$\sigma$(lys)\\
\hline
HIP 15416&	1177&	1177&	98&-&-\\		
HIP 117447&	16300&	6151&	4294&-&-\\		
HIP 2876&	3623.96&	3622&	1368&	2675&	1664\\
HIP 6522&	32600&	32600&	120620&	5488&	4345\\
HIP 3030&	5346.83&	5344&	4731&-&-\\		
HIP 100548&	40750&	5621&	4167&-&-\\		
HD 33339&	-	&-&-	&	756&	75\\
HIP 20276&	65200&	4592&	5950&	1835&	498\\
HIP 3752&	4076.95&	4075&	6571&	908&	153\\
HIP 114579&	163000&	7409&	18523&	1967&	583\\
HIP 117690&	163000&	-	&-	&21733&	57676\\
HIP 23114&	163000&	-	&-	&2243&	569\\
HIP 55892&	-&3881&	2402	&3325&	1649\\
HIP 117928&		-&3835&	3384&	3454&	1433\\
  \hline\hline
\end{tabular}
\label{tab:divin}
\begin{tablenotes}
        \footnotesize
        \item[1] Blanks in the table: data were missing in the cited references. $\sigma$ is 1 standard deviation in distance measurement.
\end{tablenotes}
\end{threeparttable}
\end{table*}

\section{Experimental configuration}

For RNG experiment, we align the telescope by pointing the telescope toward the CRSS based on its declination (DEC) and right ascension (RA) listed in the star catalogue \cite{van2007validation, masetti2008unveiling, astraatmadja2016estimating}. After we locate the CRSS and place its focal image at the center of the sensor of CCD, we employ a standard altitude-azimuth tracking mechanism to keep the focal image of the CRSS at the same position of the CCD camera as the CRSS moves \cite{kwok2016improving}, which stabilizes the coupling of photons from the CRSS into a multimode optical fiber. In the experiment, we use cosmic photons from the CRSS with wavelength in the range of [530, 680] nm for tracking, and couple photons in the range of [680, 830] nm into a multiple mode fiber for RNG.

We estimate the transmittance of the optical system to be about 20.7\% = 53\% (transmittance of RC telescope) $\times$ 75\% (transmittance of prisms) $\times$ 52\%(efficiency to couple light into multimode fiber). We estimate the transmittance of light through the atmosphere to be about 20\%. As a result, the total detection efficiency of single cosmic photon is about 2\%. (It should be noted that this efficiency is estimated for static condition and may vary, e.g., due to atmospheric disturbance.) For all of the RNG measurements in this experiment, the altitude angles of telescopes are within $70^{\circ} - 80^{\circ}$, the variation of atmospheric attenuation is insignificant according to MODTRAN model \cite{berk1987modtran}.

As shown in Fig.1, for the even-ready Bell test experiment discussed in the main text, Lab1, Lab2 and Lab for Bell state measurement (BSM) are collinear in the north-south direction (defined as the Lab axis). Lab1 has a telescope pointing toward cosmic source HIP 55892 with azimuth and altitude angles (Az1, Alt1) and optical axis at angle $\alpha 1$ with respect to the Lab axis. Lab2 has a telescope pointing toward cosmic source HIP 117928 with (Az2, Alt2) and angle $\alpha 2$. $\alpha 1$ and $\alpha 2$ are given by $\alpha = \arccos (|\cos (Alt) \cdot \cos (Az)|)$. The cosmic sources rotate around the north celestial (North Pole in the sky) as we observe on Earth (due to Earth rotation). The azimuth and altitude angles of cosmic sources change accordingly.

\begin{figure*}[tbh]
\centering
\includegraphics [width=15.6 cm,height= 9.5 cm]{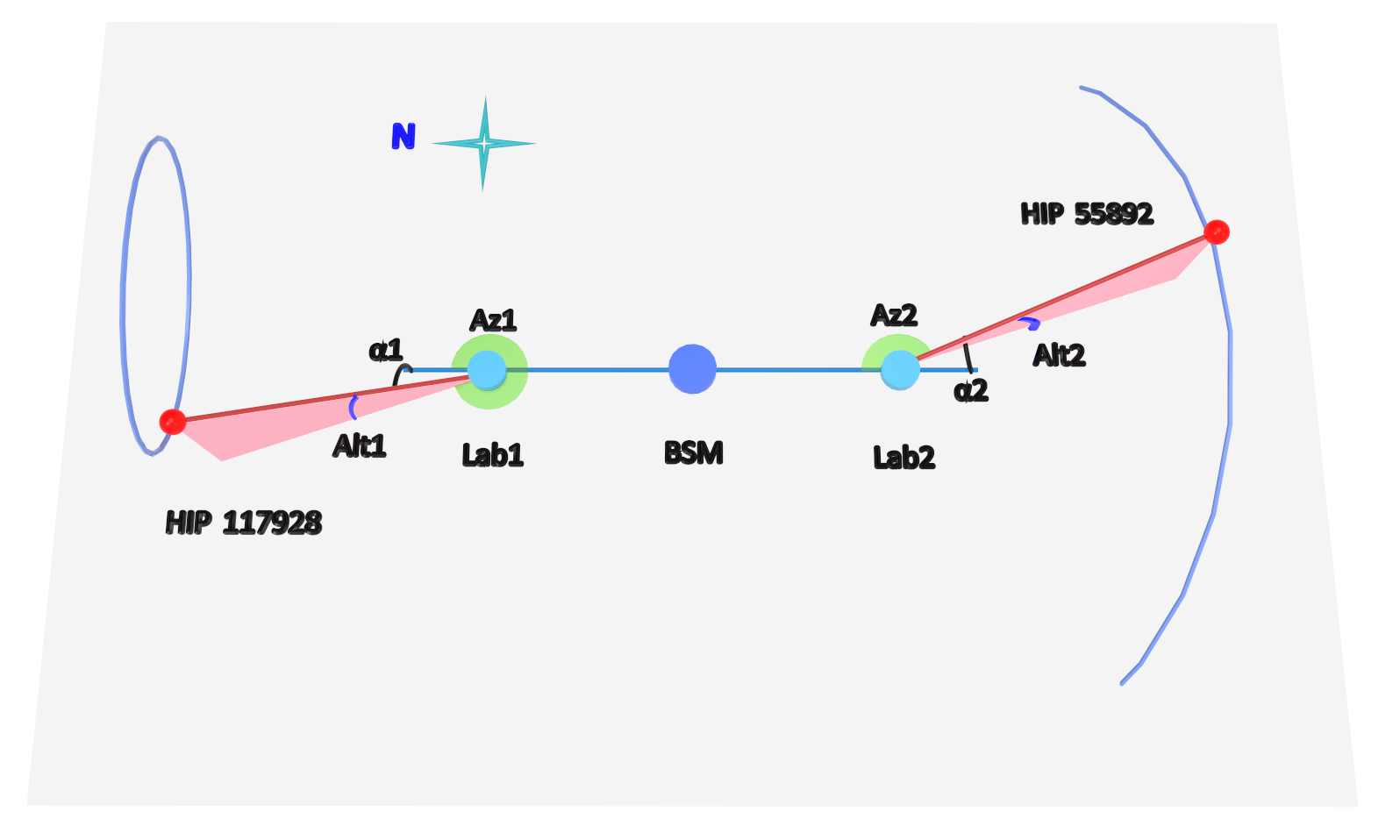}
\caption{A cartoon picture of experimental configuration. The two blue curves indicate that the cosmic sources for RNG rotate around the North Pole in the sky as we observe on Earth. For the discussion in the main text, $\alpha 1$ and $\alpha 2$ are not greater than 30 degree to satisfy space-like separation.}
\label{}
\end{figure*}

\section{Time constraints to local hidden variable mechanism}
The light cones of events in a Bell test experiment cross in the past direction, which sets time constraints to local hidden variable mechanisms, as shown in Fig. 3 in the main text.
The distances of HIP 55892 ($RA1 =171.8377^{\circ}$, $DEC1 = -24.16971 ^{\circ}$, Magnitude: 6.735) and HIP 117928 ($DEC2 = 72.46028^{\circ}$, $RA2 = 358.7932^{\circ}$, Magnitude: 8.9) from Earth are $L_1 = 3325 \pm 1649$ lys and $L_2 = 3454 \pm 1433$ lys \cite{astraatmadja2016estimating}. This gives $\tau_1 = \min (L_1/c, L_2/c)$. The angular separation $\theta$ between the two cosmic sources is given by
\begin{equation}\label{eq:1}
  \cos(\theta) = \sin(DEC1) \sin(DEC2) + \cos(DEC1) \cos(DEC2) \cos(RA1- RA2).
\end{equation}
So we have $\theta = 131.6^{\circ}$. The distance between the two cosmic sources is estimated to be $L_{12} = \sqrt{ L_1^2 + L_2^2 - 2 L_1 L_2 \cos (\theta)} = 6183$ lys. So we have $\tau_2 \geq (L_1 + L_2 + L_{12})/2c = 6481$ years, with an uncertainty of
\begin{equation*}\label{eq}
\sigma_{\tau_{2}} = \sqrt{(\sigma_{L_1}/c)^2(\tau_{2} c - L_2/2 - L_2/2 \cos (\theta))^2+ (\sigma_{L_2}/c)^2(\tau_{2} c - L_1/2 - L_1/2 \cos (\theta))^2}/(2\tau_{2} c-L_1-L_2) = 2088~years.
\end{equation*}
(The angular uncertainty is small and neglected in the estimation.)
So we have $\tau = \min ({\tau_1, \tau_2}) = 3325 \pm 1649$ years.

\section{Space-time arrangement for locality requirement}

Lab1 and Lab2 have identical experimental setups. We have $T_{Basis1} = T_{Basis2} = T_{Basis}$, $T_{Measurement1} = T_{Measurement2} = T_{Measurement}$, and $T_{w1} = T_{w2} = T_W$ for time windows (clock cycle), where $T_{Basis}$ is the time elapsed for events completing the base choice upon receiving a random bit, $T_{Measurement}$ is the time elapsed for events completing the state measurement after the base choice. We use GPS to synchronize operations such as preparations of photon-electron entanglements and reference clocks in the labs.
Locality requirement demands space-like separation between event of state measurement and event of measurement setting choice between Lab1 and Lab2,
\begin{subequations}\label{eq:2}
\begin{align}
T_{w2} + T_{Basis2} + T_{Measurement2} + L_{S1M1}/c < L_{S1M2}/c ,\\
T_{w1} + T_{Basis1} + T_{Measurement1} + L_{S2M2}/c < L_{S2M1}/c ,
\end{align}
\end{subequations}
where $L_{S1M1} (L_{S1M2})$ is distance between cosmic source and Lab1 (Lab2). Here $T_W$ is the time duration accounting for the scenario that a photon detection event occurs at the begging of a time window in Lab1 and the other one at the end of the same time window in Lab2, or vice versa. We only consider photon counting events occurring in the same time window. We then have
\begin{subequations}\label{eq:3}
\begin{align}
L_{S1M2} - L_{S1M1} \approx L_{M1M2} \cdot \cos (\alpha1)\\
L_{S2M2} - L_{S2M1} \approx L_{M1M2} \cdot \cos (\alpha2)
\end{align}
\end{subequations}
Besides, we include an additional term, $T_{Margin} \sim 1 \mu s$, which should be sufficient to account for, for example, refractive index-induced extra delay for light propagation in atmosphere (typically less than 100 nanoseconds), etc. We then have
\begin{equation}\label{eq:4}
  T_{W} + T_{Basis} + T_{Measurement} + T_{Margin} < L_{M1M2}/c \cdot \cos (\alpha),
\end{equation}
where $\alpha = \max (\alpha 1, \alpha 2)$.
Taking $T_{Measurement} \sim 4 \mu s$, $T_{Basis} < 1 \mu s$ and $\alpha < 30^{\circ}$, $T_W = 8 \mu s$, we have $L_{M1M2} > 5 km$.

\section{Space-time arrangement for freedom-of-choice requirement}

Space-like separation must be ensured between event of measurement setting choice and event of creation of photon-electron entanglement, and between event of measurement setting choice and event of BSM, so we have
\begin{subequations}\label{eq:5}
\begin{align}
L_{S1M1}/c + L_{M1B}/(c/n) - \Delta T < L_{S1B}/c,\\
L_{S2M2}/c + L_{M2B}/(c/n) - \Delta T < L_{S2B}/c,
\end{align}
\end{subequations}
where in Eq. (5a), $L_{S1M1}$ is distance between cosmic source and Lab1, $L_{S1B}$ is distance between cosmic source and BSM Lab, $L_{M1B}$ is distance between Lab1 and BSM Lab, $\Delta T$ is the time in advance to prepare photon-electron entanglement in Lab1 (see Fig.3), and n = 1.45 is the refractive index of fiber to connect Lab1 and BSM lab; Eq. (5b) describes the same conditions in Lab2.
With BSM lab in the middle and identical conditions at both sides, we have
\begin{equation}\label{eq:6}
\Delta T > (1.45-\cos \alpha) \cdot L_{M1M2} / 2c \sim 5 \mu s.
\end{equation}
Or, we have $\Delta T + T_{W} + T_{Basis} +T_{Margin} \sim$ 15 $\mu s$ before state measurement, which is much shorter than the coherence time (T2 $\sim 0.6 s$ at 77 K \cite{bar2013solid}) of electron spin of NV center.

\section{Estimation about event-ready Bell test experiment with cosmic RNGs }

The altitude angles of telescopes pointing toward HIP 55892 ($171.8377^{\circ}$, $-24.16971^{\circ}$) and HIP 117928 ($72.46028^{\circ}$, $358.7932^{\circ}$) change due to Earth rotation. For the event-ready Bell experiment discussed in the main text, there is a limited time window ($\sim 2 hours$) each night to collect photons from the two cosmic sources simultaneously. For example, from 3 AM to 5 AM on January 13th, 2017, the azimuth and altitude angles of HIP 55892 change from ($162^{\circ}$, $23^{\circ}$) to ($192^{\circ}$, $25^{\circ}$), while the azimuth and altitude angles of HIP 117928 change from ($352^{\circ}$, $24^{\circ}$) to ($2^{\circ}$, $23^{\circ}$). Correspondingly, $\alpha 1$ changes from $29^{\circ}$ to $28^{\circ}$, and $\alpha 2$ changes from $25^{\circ}$ to $23^{\circ}$.

The average signal rate for altitude $80^{\circ}$ and cosmic sources with magnitude $\sim 6.7$ is $r (80^{\circ}$, 6.7 mag) = 510,000 $s^{-1}$, similarly, we have $r (80^{\circ}$, 8.9 mag) = 107,000 $s^{-1}$ as shown in Fig. 2 in the main text. We use MODTRAN model \cite{berk1987modtran} to scale the signal rates to $r1 (23^{\circ}$, 6.7 mag) = 324,000 $s^{-1}$ (with signal-to-noise ration of 590) and $r2 (23^{\circ}$, 8.9 mag) = 68,000 $s^{-1}$ (with signal-to-noise ration of 124). By having $T_W = 8 \mu s$, we estimate the probability for the two RNGs to output a pair of random bits in the same time window to be $P_{RNG} = (1 - e^{-r1 \cdot T_W})(1 - e^{-r2 \cdot T_W}) \sim 0.39$.

The photons emitted by NV centers at the visible wavelength ($\sim$ 640 nm) are downconverted to photons at the wavelength of $\sim$ 1550 nm via difference frequency generation (DFG), because they are subject to high propagation loss in the optical fiber. Considering the 1.5 dB of photon loss for downconversion operation \cite{de2012quantum} and 1 dB of photon loss for photon propagation over the 5 km optical fiber, that will be an improvement of $\sim$ 10 dB in overall two-photon detection efficiency over the previous experiment \cite{hensen2015loophole}. With that, the averaged success probability per entanglement generation attempt is estimated to be $P_{total} = 2.26 \times 10^{-8}$. This will result in one event-ready entanglement per 0.29 hours per 24 $\mu s$ measurement period on average, and about 72 hours to violate the Bell's inequality with statistical confidence similar to what was achieved by Hensen et al. \cite{hensen2015loophole}, but with an improvement of setting the time constraint to be by more than 1000 years before the experiment.

\section{Estimating signal-to-noise ratio for RNG with photons from quasar}
The background rate in the current experiment is 550 $s^{-1}$. The Ritchey-Chretien (RC) telescope used in the current experiment has a diameter $d = 1$ meter and a focal length $f = 5$ meter. The beam diameter at waist and field-of-view are $10.2 \mu m$ and 0.3 arcsec$^2$ in the diffraction limit, while they are estimated to be 73 $\mu m$ and 14.7 arcsec$^2$ in the current experiment. It has been demonstrated that the optical telescope can be made to work in the diffraction limit in previous astronomy studies \cite{wizinowich2000first}. Assuming uniformity for sky brightness, the background rate is proportional to the field-of-view. By having the RC telescope work in the diffraction limit, the background rate will be reduced to $\frac{0.3}{14.7}\times 550 = 11.2$ $s^{-1}$. The true signal rate for RNG with photons from quasar APM 08279+5255 is 800 $s^{-1}$ at $\alpha = 80^\circ$ based on the trend line of Fig. 2 in the main text, and 590 $ s^{-1}$ at $\alpha = 30^\circ$  based on MODTRAN model. So the signal-to-noise ratio is 53.7.

Although the reported sky brightness at Xinglong is in the range of [16, 22] magnitude per arcsec$^{2}$ \cite{zhang2015astronomical}, which is similar to what was observed in other observatories. Being located at a lower altitude and close to major cities (Beijing, Chengde, and a few others), the experiment at Xinglong observatory may be affected by scattered light due to human activities. The sky brightness can be reduced when working at higher altitude and being away from human activities.

By performing the same measurement in the space, there is no atmosphere attenuation and the sky brightness is reduced, the signal-to-noise ratio can be increased to $>$ 100. (We estimate the atmospheric transmittance to be $\sim$ 20\% in the current measurement.) In addition, the absence of atmospheric disturbance will ease the task to optimize the optical system. The two telescopes can be set with angular separation to be $180^{\circ}$, allowing to access more celestial objects in the universe. Overall, a satellite-based cosmic Bell test experiment may have better performance.

\section{Probability Distribution of photon detection events versus time bins}

The raw data from our random number generator is very close to a uniform distribution, as shown in Fig. 2. The probability for a photon detected in each of $N_b$ bins is $P = 1/N_b = 1/256$.

\begin{figure*}[tbh]
\centering
\includegraphics [width=18.1cm,height=12.6cm]{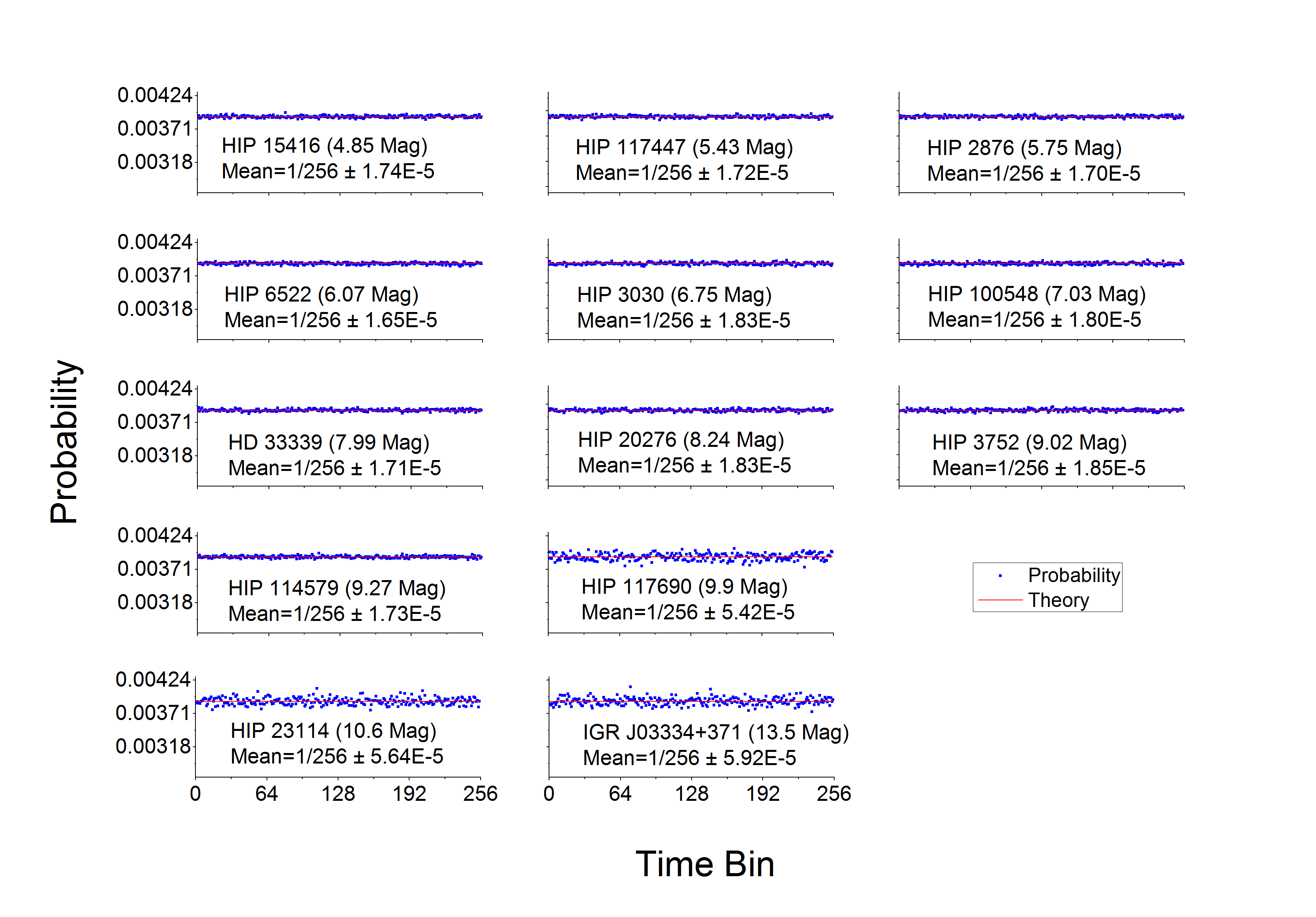}
\caption{Experimental probability distribution (scatter dots) of photon-detection events versus time bin. Smooth line (red) is for theoretical value = 1/256.}
\label{}
\end{figure*}

\section{NIST Test Result}

We employ the NIST statistical test suite \cite{nist2010special} to assess randomness of the raw data in cosmic photons measurements. The NIST test results are shown in the Table below.

\begin{table*}[htb]
\caption{NIST test results for random bits. The worst outcomes are selected from the tests that produce multiple outcomes of p-values and proportions. }
\centering
\begin{tabular}{ccccccc}
\hline\hline
Name(Vmag)&	\multicolumn {2}{c} {HIP 15416(4.85)}&\multicolumn {2}{c} {HIP 117447(5.43)}&\multicolumn {2}{c} {HIP 2876(5.75)}\\
STATISTICAL TEST&   P-VALUE&	PROPORTION&	P-VALUE&	PROPORTION&	P-VALUE&	PROPORTION\\
\hline
Frequency& 0.585209& 990/1000& 0.936823& 990/1000& 0.639202& 992/1000 \\
BlockFrequency& 0.008753& 991/1000& 0.859637& 992/1000& 0.842937& 996/1000 \\
CumulativeSums& 0.036113& 994/1000& 0.672470& 993/1000& 0.450297& 991/1000 \\
Runs& 0.118120& 991/1000& 0.262249& 988/1000& 0.548314& 988/1000 \\
LongestRun& 0.289667& 995/1000& 0.585209& 991/1000& 0.439122& 986/1000 \\
Rank& 0.329850& 991/1000& 0.632955& 991/1000& 0.593478& 987/1000 \\
FFT& 0.858002& 990/1000& 0.360287& 990/1000& 0.643366& 989/1000 \\
NonOverlappingTemplate& 0.007057& 986/1000& 0.008691& 993/1000& 0.004301& 988/1000 \\
OverlappingTemplate& 0.729870& 984/1000& 0.171867& 990/1000& 0.037320& 992/1000 \\
Universal& 0.270265& 997/1000& 0.319084& 984/1000& 0.278461& 985/1000 \\
ApproximateEntropy& 0.603841& 991/1000& 0.031219& 984/1000& 0.059358& 990/1000 \\
RandomExcursions& 0.114131& 588/594& 0.207637& 585/592& 0.132687& 635/642 \\
RandomExcursionsVariant& 0.007675& 589/594& 0.037192& 583/592& 0.125699& 636/642 \\
Serial& 0.444691& 987/1000& 0.274341& 991/1000& 0.142872& 984/1000 \\
LinearComplexity& 0.353733& 991/1000& 0.353733& 988/1000& 0.304126& 991/1000 \\

  \hline\hline
\end{tabular}
\label{tab:divin}
\end{table*}

\begin{table*}[htb]
\centering
\begin{tabular}{cccccccc}

  \hline\hline
  Name(Vmag)&	\multicolumn {2}{c} {HIP 6522(6.07)}&\multicolumn{2}{c}{HIP 3030(6.75)}&\multicolumn {2}{c} {HIP 100548(7.03)}&\\
  STATISTICAL TEST&	P-VALUE&	PROPORTION&	P-VALUE&	PROPORTION&	P-VALUE&	PROPORTION\\
  \hline
Frequency& 0.045971& 989/1000& 0.012300& 993/1000& 0.945296& 986/1000 \\
BlockFrequency& 0.264901& 988/1000& 0.461612& 991/1000& 0.522100& 993/1000 \\
CumulativeSums& 0.169981& 991/1000& 0.783019& 995/1000& 0.134172& 990/1000 \\
Runs& 0.467322& 995/1000& 0.632955& 987/1000& 0.837781& 992/1000 \\
LongestRun& 0.238035& 986/1000& 0.028817& 983/1000& 0.189625& 985/1000 \\
Rank& 0.148653& 989/1000& 0.936823& 988/1000& 0.630872& 991/1000 \\
FFT& 0.123755& 991/1000& 0.083526& 989/1000& 0.016833& 987/1000 \\
NonOverlappingTemplate& 0.009810& 985/1000& 0.008385& 989/1000& 0.003273& 985/1000 \\
OverlappingTemplate& 0.112047& 984/1000& 0.686955& 987/1000& 0.350485& 986/1000 \\
Universal& 0.206629& 993/1000& 0.069863& 989/1000& 0.222480& 986/1000 \\
ApproximateEntropy& 0.018036& 993/1000& 0.005054& 988/1000& 0.807412& 990/1000 \\
RandomExcursions& 0.027252& 597/608& 0.413802& 609/620& 0.195163& 603/608 \\
RandomExcursionsVariant& 0.040547& 594/608& 0.048716& 612/620& 0.029140& 601/608 \\
Serial& 0.262249& 989/1000& 0.152902& 991/1000& 0.788728& 992/1000 \\
LinearComplexity& 0.784927& 990/1000& 0.939005& 989/1000& 0.595549& 990/1000 \\

  \hline\hline
  \end{tabular}
  \label{tab:divin}
  \end{table*}

 \begin{table*}[htb]
\centering
\begin{tabular}{ccccccc}

  \hline\hline
  Name(Vmag)&\multicolumn {2}{c} {HD 33339(7.99)}&	\multicolumn {2}{c} {HIP 20276(8.24)}&	\multicolumn {2}{c} {HIP 3752(9.02)}\\
  STATISTICAL TEST&	P-VALUE&	PROPORTION&	P-VALUE&	PROPORTION&	P-VALUE&	PROPORTION\\
  \hline
Frequency& 0.945296& 994/1000& 0.851383& 997/1000& 0.997292& 992/1000 \\
BlockFrequency& 0.145326& 987/1000& 0.072066& 990/1000& 0.234373& 985/1000 \\
CumulativeSums& 0.279844& 994/1000& 0.488534& 998/1000& 0.676615& 994/1000 \\
Runs& 0.691081& 990/1000& 0.116746& 990/1000& 0.077131& 990/1000 \\
LongestRun& 0.983938& 986/1000& 0.214439& 991/1000& 0.135720& 990/1000 \\
Rank& 0.380407& 991/1000& 0.841226& 994/1000& 0.457825& 990/1000 \\
FFT& 0.007264& 989/1000& 0.574903& 990/1000& 0.455937& 991/1000 \\
NonOverlappingTemplate& 0.007212& 994/1000& 0.007160& 988/1000& 0.024028& 988/1000 \\
OverlappingTemplate& 0.402962& 994/1000& 0.317565& 987/1000& 0.792508& 994/1000 \\
Universal& 0.048404& 987/1000& 0.862883& 991/1000& 0.917870& 990/1000 \\
ApproximateEntropy& 0.474986& 990/1000& 0.109435& 988/1000& 0.240501& 987/1000 \\
RandomExcursions& 0.021682& 604/612& 0.020233& 617/624& 0.009706& 598/607 \\
RandomExcursionsVariant& 0.022671& 609/612& 0.136777& 621/624& 0.206812& 603/607 \\
Serial& 0.102526& 988/1000& 0.152044& 991/1000& 0.100709& 990/1000 \\
LinearComplexity& 0.771469& 991/1000& 0.662091& 983/1000& 0.940080& 988/1000 \\

  \hline\hline
  \end{tabular}
  \label{tab:divin}
  \end{table*}

\begin{table*}[htb]
    \centering
    \begin{tabular}{ccccccc}
    \hline\hline
  Name(Vmag) &		\multicolumn {2}{c} {HIP 114579(9.27)}&	\multicolumn {2}{c} {HIP 117690(9.9)} & \multicolumn {2}{c} {HIP 23114(10.6)}\\
  STATISTICAL TEST&    P-VALUE&	PROPORTION&	P-VALUE&	PROPORTION&	P-VALUE&	PROPORTION\\
  \hline
Frequency& 0.142062& 989/1000& 0.911413& 98/100& 0.419021& 100/100 \\
BlockFrequency& 0.293952& 994/1000& 0.366918& 99/100& 0.085587& 98/100 \\
CumulativeSums& 0.528111& 995/1000& 0.514124& 98/100& 0.181557& 100/100 \\
Runs& 0.328297& 989/1000& 0.924076& 99/100& 0.437274& 99/100 \\
LongestRun& 0.025193& 990/1000& 0.554420& 100/100& 0.191687& 99/100 \\
Rank& 0.899171& 986/1000& 0.122325& 100/100& 0.224821& 100/100 \\
FFT& 0.016149& 989/1000& 0.816537& 99/100& 0.574903& 99/100 \\
NonOverlappingTemplate& 0.003297& 986/1000& 0.013569& 98/100& 0.008266& 98/100 \\
OverlappingTemplate& 0.013102& 991/1000& 0.616305& 97/100& 0.554420& 98/100 \\
Universal& 0.641284& 989/1000& 0.383827& 100/100& 0.514124& 98/100 \\
ApproximateEntropy& 0.010531& 985/1000& 0.897763& 100/100& 0.191687& 99/100 \\
RandomExcursions& 0.000508& 636/640& 0.031497& 64/64& 0.070445& 67/67 \\
RandomExcursionsVariant& 0.020592& 628/640& 0.060239& 63/64& 0.011931& 67/67 \\
Serial& 0.242986& 991/1000& 0.401199& 99/100& 0.514124& 98/100 \\
LinearComplexity& 0.693142& 993/1000& 0.657933& 99/100& 0.122325& 99/100 \\

  \hline\hline
\end{tabular}
\label{tab:divin}
\end{table*}

(The data for quasar fails the NIST test, partly due to a small statistics and relatively large background contribution.)

%

\bibliographystyle{apsrev4-1}
